\documentclass[aip, amsmath,amssymb,reprint]{revtex4-1}

\usepackage{diagbox}

\usepackage{graphicx}% Include figure files
\usepackage{dcolumn}% Align table columns on decimal point
\usepackage{bm}% bold math

\usepackage[utf8]{inputenc}
\usepackage[T1]{fontenc}
\usepackage{mathptmx}
\usepackage{etoolbox}

\usepackage{amssymb,amsmath,placeins,epsfig,array}
\usepackage[usenames,dvipsnames]{color}
\usepackage[normalem]{ulem}
\usepackage[breaklinks,colorlinks,urlcolor=blue,citecolor=blue,linkcolor=blue]{hyperref}
\usepackage{lineno}
\usepackage{graphicx}
\usepackage{float}
\usepackage{mathptmx}

\usepackage{scalerel}
\usepackage{tikz}
\usetikzlibrary{svg.path}

\definecolor{orcidlogocol}{HTML}{A6CE39}
\tikzset{
  orcidlogo/.pic={
    \fill[orcidlogocol] svg{M256,128c0,70.7-57.3,128-128,128C57.3,256,0,198.7,0,128C0,57.3,57.3,0,128,0C198.7,0,256,57.3,256,128z};
    \fill[white] svg{M86.3,186.2H70.9V79.1h15.4v48.4V186.2z}
                 svg{M108.9,79.1h41.6c39.6,0,57,28.3,57,53.6c0,27.5-21.5,53.6-56.8,53.6h-41.8V79.1z M124.3,172.4h24.5c34.9,0,42.9-26.5,42.9-39.7c0-21.5-13.7-39.7-43.7-39.7h-23.7V172.4z}
                 svg{M88.7,56.8c0,5.5-4.5,10.1-10.1,10.1c-5.6,0-10.1-4.6-10.1-10.1c0-5.6,4.5-10.1,10.1-10.1C84.2,46.7,88.7,51.3,88.7,56.8z};
  }
}

\newcommand\orcidicon[1]{\href{https://orcid.org/#1}{\mbox{\scalerel*{
\begin{tikzpicture}[yscale=-1,transform shape]
\pic{orcidlogo};
\end{tikzpicture}
}{|}}}}

\usepackage{hyperref}

\usepackage{graphicx}% Include figure files
\usepackage{dcolumn}% Align table columns on decimal point
\usepackage{bm}% bold math
\usepackage{subfigure}
\usepackage{epstopdf}
\usepackage{hyperref}
\usepackage{color}
\usepackage[utf8]{inputenc}

\makeatletter
\def\@email#1#2{%
 \endgroup
 \patchcmd{\titleblock@produce}
  {\frontmatter@RRAPformat}
  {\frontmatter@RRAPformat{\produce@RRAP{*#1\href{mailto:#2}{#2}}}\frontmatter@RRAPformat}
  {}{}
}%
\makeatother
\begin{document}

\preprint{AIP/123-QED}
\title{\large{Adiabatic Light Guide with S-shaped Strips}}

\author {B.~Wojtsekhowski \orcidicon{0000-0002-2160-9814}} 
\email[Contact person, ]{bogdanw@jlab.org} 
\affiliation{Physics Division, Thomas Jefferson National Accelerator Facility{,} Newport  News{,} VA 23606, USA} 
\author {E.J.~Brash}
\affiliation{\mbox{Department of Physics, Christopher Newport University, Newport News, VA 23606, USA}}
\affiliation{Physics Division, Thomas Jefferson National Accelerator Facility{,} Newport  News{,} VA 23606, USA} 
\author {G.B.~Franklin}
\affiliation{\mbox{Department of Physics, Carnegie Mellon University, Pittsburgh, PA 15213, USA}}
\author {A.~Rosso}
\affiliation{\mbox{Department of Physics, Christopher Newport University, Newport News, VA 23606, USA}}
\author {A.~Sarty}
\affiliation{\mbox{Department of Astronomy and Physics, Saint Mary's University, Halifax, NS B3H 3C3, Canada}}
\author {A.~Shahinyan}
\affiliation {\mbox{Experimental Physics Division, A.I.~Alikhanyan National Science Laboratory, Yerevan, 0036, Armenia}}
\author {E.~Zimmerman}
\address {\mbox{Department of Physics, James Madison University, Harrisonburg, VA 22807, USA}}
 
\begin{abstract}
A light guide is an essential part of many scintillator counters and light collection systems. 
Our main interest is a light guide for a thin wide scintillator which has high light transmission while converting the area of the light source to the shape of a photo-detector.
We propose a variation of the light guide which avoids a 90$\rm ^o$ twist of the strips, reduces the length of the light pipe, and reduces the complexity of production.
Detailed Monte Carlo simulation studies have been performed for a 3-strip S-shaped light-guide system.
\end{abstract}

\date{\today}

\maketitle

%\linenumbers

\section{Introduction}

A light guide for scintillator counters has been the subject of many investigations since the 1950s, see Ref.~\cite{Garwin},
and continues with technological advances, see e.g. Refs.~\cite{Dougan, Cheng, Olivenboim}.
The applicability of Liouville's theorem to the light propagation in geometrical optics means invariance of the phase space $\Delta x \times \Delta \theta_x$ in the x-direction (and similarly in the y-direction) when the cross-section area of the light guide changes sufficiently slowly.
Such a configuration is often called an ``adiabatic" light guide.
As a result, the photons produced in the scintillator can propagate without loss, allowing coupling of the scintillator counter to the photo detector, typically a vacuum photomultiplier tube (PMT)~\cite{Wright, Leo}.

\begin{figure}[ht!]
	\centering
        \includegraphics[width=0.6 \linewidth, angle = 90.]{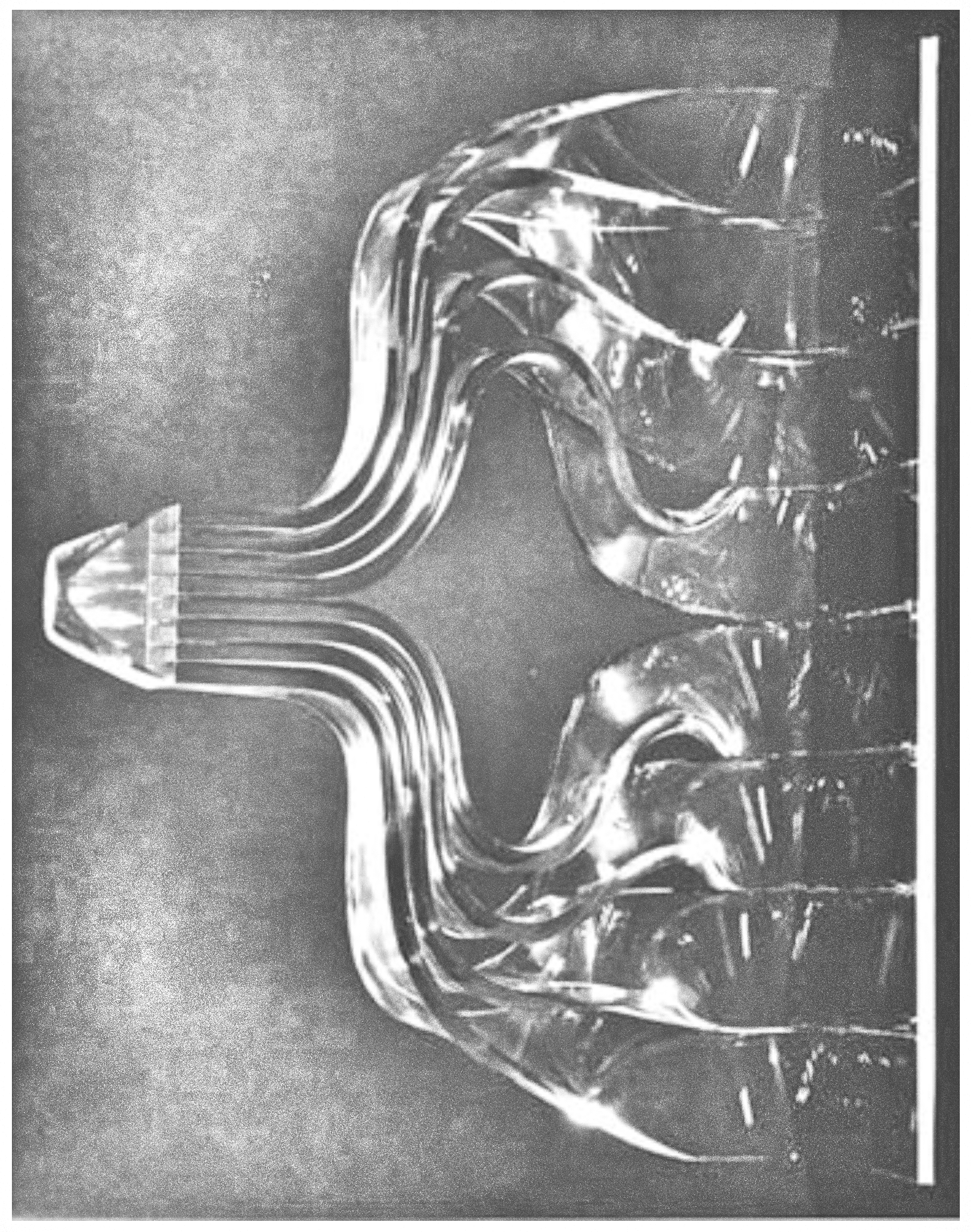}
	\caption{Classical adiabatic light guide with twisted strips. The PMT is attached to the LG through the conical section. 
    The scintillator is attached to the top side of the LG.This picture is taken from Ref.~\cite{Grupen}. 
    Reproduced with permission from Particle detectors, (2008). Copyright 2008 Cambridge University Press.}
\label{fig:ALG_old}
\end{figure}

A light guide (LG) for a wide and thin scintillator counter was proposed in Ref.~\cite{Gorenstein} by means of several constant cross section strips (machined as straight) which are slowly twisted by 90$\rm ^o$, see also Refs.~\cite{Dougan, Cheng, Grupen}. 
An example of such an LG is shown in Fig.~\ref{fig:ALG_old}.
Near the PMT, where the number of the remaining photons' bounces is small, a conical section was used for further concentration of the light. 

The standard adiabatic LG~\cite{Gorenstein} with a 90$\rm ^o$ rotation of the strips involves labor consuming construction.
The rotation needs to be done gradually, which means the rotation angle is much less than one radian per length of the LG equal to the largest transverse dimension of the strip. 
This leads to a significant length of the LG.
Currently, there are several companies able to produce twisted light guides; see information on the web pages~\cite{web:ELJEN, web:Epic, web:Luxium}.

\begin{figure}[ht!]
	\centering
        \includegraphics[angle = 0, width=0.8 \linewidth]{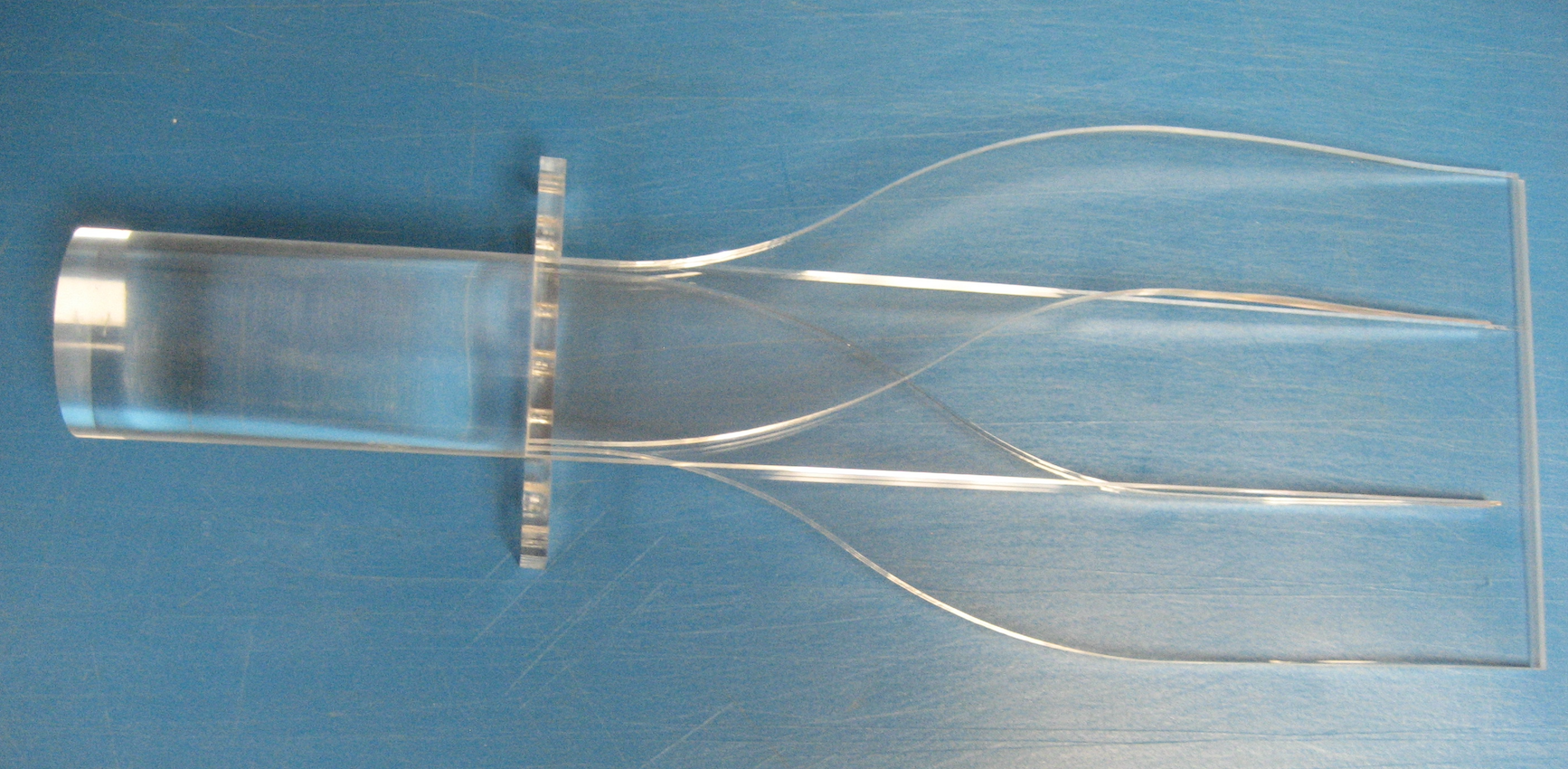}
	\caption{Light guide made of three strips: two S-shaped and a straight central one.}
\label{fig:ALG_3}
\end{figure}

We are proposing a novel geometry adiabatic LG in which the 90$\rm ^o$ twist is avoided.
Instead of twisted strips, our LG is made of flat machined S-shaped strips which are modified by a modest out-of-plane bend, as shown in Fig.~\ref{fig:ALG_3}.
Inexpensive production of these strips has become possible due to the availability of modern milling machines, and also can be done by using laser cutting~\cite{Bahr, Mamyan}.
A large number of such LGs have been constructed at Thomas Jefferson National Accelerator Facility (JLab) and Carnegie Mellon University (CMU); see, for example, Refs.~\cite{HallA_2007, HCAL}.

The proposed LG concept could also be useful in solar energy systems as it allows the transfer of the light collected in a large area to a compact photo detector (for example an LG between a thin wave-length shifter to a silicon-based diode).
We anticipate that this paper will lead to the realization of the novel LG in many modern scintillator detectors.

Monte Carlo simulations (MC) of our LG concept were previously performed using Guide-7~\cite{Glister} and Geant4~\cite{Mamyan}.
In the current report, we present the results of our simulations made using Geant4, a Monte-Carlo based toolkit for simulating nuclear physics which is sourced in C++~\cite{agostinelli_2003}. 
Currently, generic Geant4 has all the components needed for the LG simulation.
A three-strip example is shown in Fig.~\ref{fig:ALG_3}.
The performance of the LG with five and seven strips is also investigated.

We have added a description of the custom parts of the code (such as the twisted LG and the S-shaped strip) in an appendix to this paper for the convenience of the reader.
Internally consistent and reproducible results were obtained in the simulations by specifying one thread per CPU.
Refs.~\cite{Zimmerman, Rosso} have the source codes of our MC.

\section{Optics of the photon in light guide}
\label{sec:fresnel}
In designing the specific LG it is useful to get quantitative estimates of light propagation efficiency, so we included in this section a discussion of the LG basics for exactly such a reason.

Modern MC tools allow us to study LG performance and to do optimization of light propagation.
The practically important parameter of LG is the fraction of the photons transmitted to the PMT.
Fresnel's equations used in Geant4 provide a full description of the photon interaction with a smooth surface.
Imperfections in the surface were not included in our MC.

\begin{figure}[ht!]
	\centering
        \includegraphics[width=0.5 \linewidth]{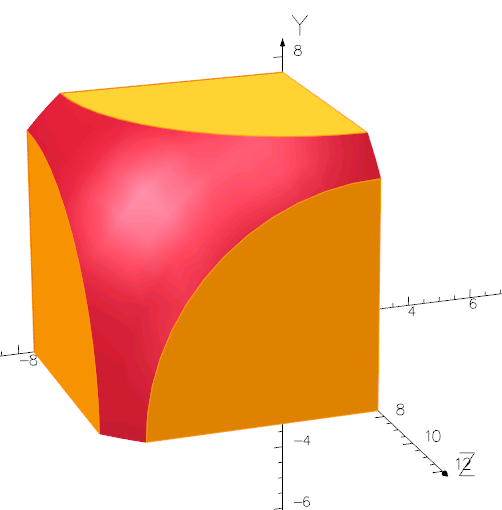}
	\caption{Diagram illustrating the optical properties of a right angle block (1/8 of the phase space is shown).}
\label{fig:cone}
\end{figure}
Photons with an incident trajectory between a normal to the surface and a critical angle, $\theta \, =$ 41.8$\rm ^o$ (for a plastic with refraction index n = 1.50), are able to escape the scintillator (yellow cones in Fig.~\ref{fig:cone}).
When the photons have an inclination angle to the block surface above 41.8$\rm ^o$, the photons are internally reflected inside the block (red area in Fig.~\ref{fig:cone}) until they reach the LG after a large number of reflections.

For isotropic photon emission by the scintillator material, the transmission fraction on one side of the scintillator equals $(1-\cos{\theta})/2$ or 0.123 (out of $4\pi$).
With one PMT in the scintillator counter, the photons on the PMT side will be detected, but the photons escape detection when emitted in the other five directions.
The remaining fraction of 0.262 (shown in red in Fig.~\ref{fig:cone}) will reach the LG after a large number of reflections.
The actual fraction of photons reaching the PMT could be found via MC simulation.

\section{Light transmission by a bent strip}

An MC study has been done for a single strip bent out-of-plane and in-plane.
As a photon source an isotropic emitter was used, located in the middle of a straight section (scintillator) before the LG. 
We counted photons in the PMT attached to the end of the LG and calculated the transmission coefficient. 
The MC simulations presented here have 50k initial photons.

\begin{figure}[ht!]
	\centering
        \includegraphics[trim = 100 50 100 10, width= 0.45 \linewidth]{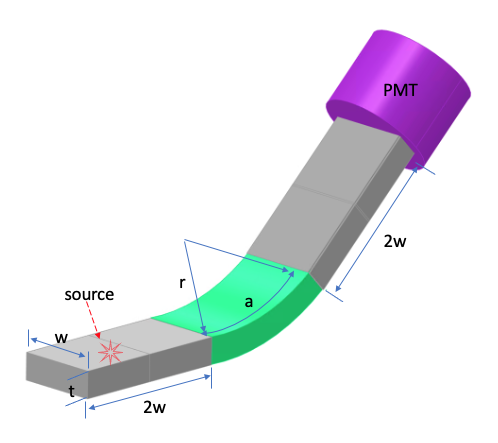}
	\caption{The element of geometry of the out-of-plane bent strip.
 The curved section of the strip is shown in green.
 The photon source (shown as a red star) is located in the middle of the left straight section.}
\label{fig:3D-out}
\end{figure}

The out-of-plane case, shown in Fig.~\ref{fig:3D-out}, is characterized by the ratio of the bend radius, $\bf r$, to the thickness of the strip in the direction of the bend, {$\bf t$}, and the bending angle, $\bf a/r$, where $\bf a$ is the length of the curved area.
The thickness of the strip, $\bf t$, is 0.5 cm.
The width of the sections, $\bf w$, is 5 cm.
\vskip -0.35 in
\begin{table}[!htb]
    \centering
    \caption{The results of MC simulation of the transmission probability through a geometry shown in Fig.~\ref{fig:3D-out} for the out-of-plane bent strip vs. the geometrical parameters $\bf r/t$ and $\bf a/r$.}
    \begin{tabular}{|c |p{0.09 \linewidth} |p{0.09 \linewidth}|p{0.09 \linewidth}|p{0.09 \linewidth}|p{0.09 \linewidth}|p{0.09 \linewidth}|  }
        \hline
        \backslashbox{\textbf{a\,/\,r [rad]}}{\textbf{r \,/\, t}} & 100 & 50 & 20 & 10 & 5 &1 \\
        \hline
        1.0 &98.1 &97.8 &96.6 &94.6 &90.7 &76.7 \\  
        0.5 &98.1 &97.8 &96.7 &94.9 &91.4 &83.8 \\  
        0.2 &98.2 &97.9 &97.0 &95.6 &93.1 &90.1 \\  
        0.1 &98.2 &98.0 &97.3 &96.0 &94.8 &93.8 \\ \hline    
    \end{tabular}
    \label{tab:in-bend}
\end{table}

\begin{figure}[ht!]
	\centering
        \includegraphics[trim = 100 0 100 0, width= 0.45 \linewidth]{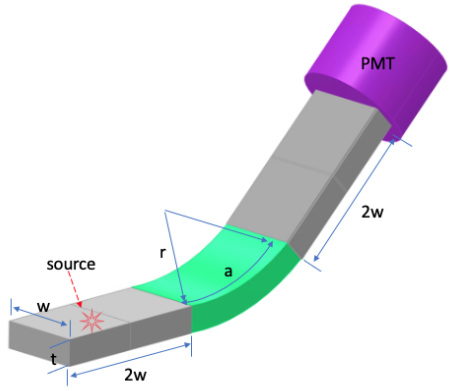}
	\caption{The geometry of the in-plane bent strip. 
    See also the caption for Fig.~\ref{fig:3D-out}.}
\label{fig:3D-in}
\end{figure}

As one can see in Tab.~\ref{tab:in-bend}, the transmission is still high (91\%), even for a relatively sharp bend $\bf r/t = 5$ and a one radian bend angle, which is not obvious from the concept of an adiabatic LG.
The in-plane case, shown in Fig.~\ref{fig:3D-in}, is also characterized by the ratio of the bend radius, $\bf r$, to the size of the strip in the direction of the bend, $\bf w$, and the bending angle, $\bf a/r$.

\begin{table}[!ht]
    \centering
    \caption{The results of MC simulation of the transmission probability through a light guide shown in Fig.~\ref{fig:3D-in} for the in-plane bending LG vs. the geometrical parameters $\bf r/w$ and $\bf a/r$.}
    \begin{tabular}{|c |p{0.09 \linewidth} |p{0.09 \linewidth}|p{0.09 \linewidth}|p{0.09 \linewidth}|p{0.1 \linewidth}|p{0.09 \linewidth}|  }
        % \hline
        % & \multicolumn{6}{c|}{\textbf{r \,\,/\,\, w}} \\
        \hline 
        \backslashbox{\textbf{a\,/\,r [rad]}}{\textbf{r \,/\, w}} & 100 & 50 & 20 & 10 & 5 &1 \\
        \hline
        1.0 &98.4 &98.1 &96.9 &95.1 &90.2 &76.4 \\  
        0.5 &98.4 &98.1 &97.0 &95.3 &91.5 &82.5 \\  
        0.2 &98.5 &98.1 &97.3 &96.0 &93.6 &90.1 \\  
        0.1 &98.5 &98.2 &97.7 &96.7 &95.2 &94.3 \\
        \hline
    \end{tabular}
    \label{tab:out-bend}
\end{table}

The results for the in-plane bend are very similar to the case of the out-of-plane bend:
The transmission is still high (90\%) even for a relatively sharp bend $\bf r/t = 5$ and a one radian bend angle.

\begin{figure}[!ht]
	\centering
        \includegraphics[trim = 150 0 0 150, angle = 0, width= 0.7 \linewidth]{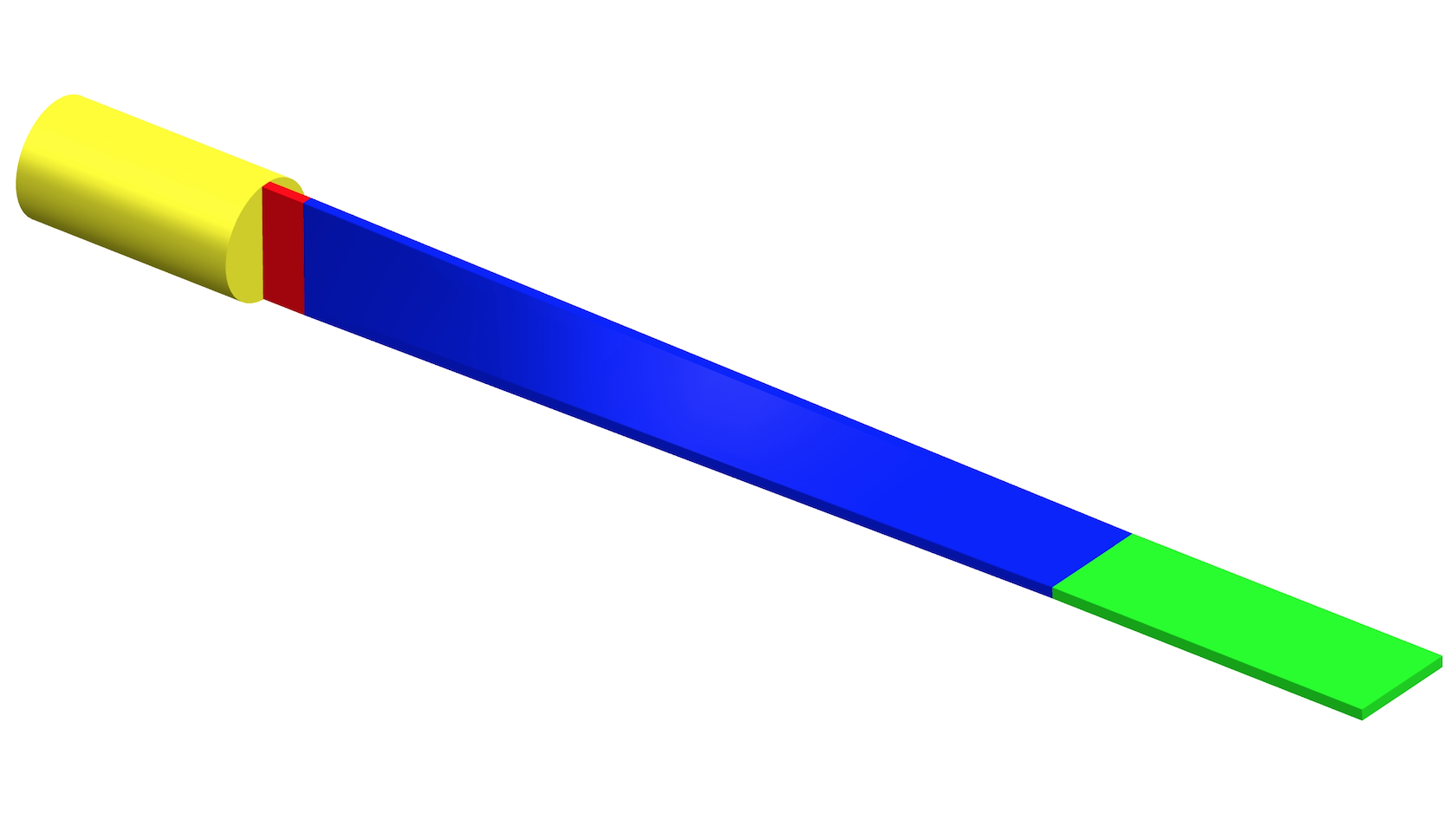}
	\caption{The view of the 90$^o$ twisted strip.}
\label{fig:twisted-strip}
\end{figure}

Light propagation was also investigated for the twisted strip shown
in Fig.~\ref{fig:twisted-strip}.
The width of the strip is $\bf w$ = 5~cm and the full length is $\bf l$.
The total rotation angle of the strip plane is $\bf \theta$.
The thickness of the strip was 0.5~cm, as in the other investigated LGs.
 
\begin{table}[!ht]
    \centering
    \caption{The results of MC simulation of the transmission probability through a twisted strip shown in Fig.~\ref{fig:twisted-strip} for different twist angle.}
    \begin{tabular}{|c |p{0.09 \linewidth} |p{0.09 \linewidth}|p{0.09 \linewidth}|p{0.09 \linewidth}|p{0.09 \linewidth}|p{0.09 \linewidth}|  }
        \hline 
        \backslashbox{\textbf{\,$\theta$ [rad]}}{\textbf{l \,/\,w}} & 100 & 50 & 20 & 10 & 5 &1 \\
        \hline
        $\pi/2$ &96.3 &94.1 &87.7 &79.1 &66.0 &28.5 \\  
        $\pi/3$ &96.4 &94.3 &88.6 &80.3 &67.9 &32.1 \\  
        $\pi/4$ &96.2 &94.5 &88.9 &81.8 &69.0 &34.3 \\  
        $\pi/6$ &96.3 &94.6 &89.1 &82.6 &70.8 &37.7 \\
        \hline
    \end{tabular}
    \label{tab:twist-1}
\end{table}

As shown in Tab.~\ref{tab:twist-1} for a 90$^o$ twist of the strip the light propagation probability is about 90\% for 
$\bf l/w \sim$25 or $\bf l$ = 125~cm.
Comparison of the last table with the in-plane and out-of-plane bends (Tabs.~\ref{tab:in-bend}, \ref{tab:out-bend}) indicates a much larger loss for the combined bend case at the same length of the LG.
\section{The photon transmission in the 90$^o$ twist LG and the proposed S-shaped LG}

\begin{figure}[!ht]
         \includegraphics[trim = 0 90 0 90, angle = 0, width= 0.65 \linewidth]{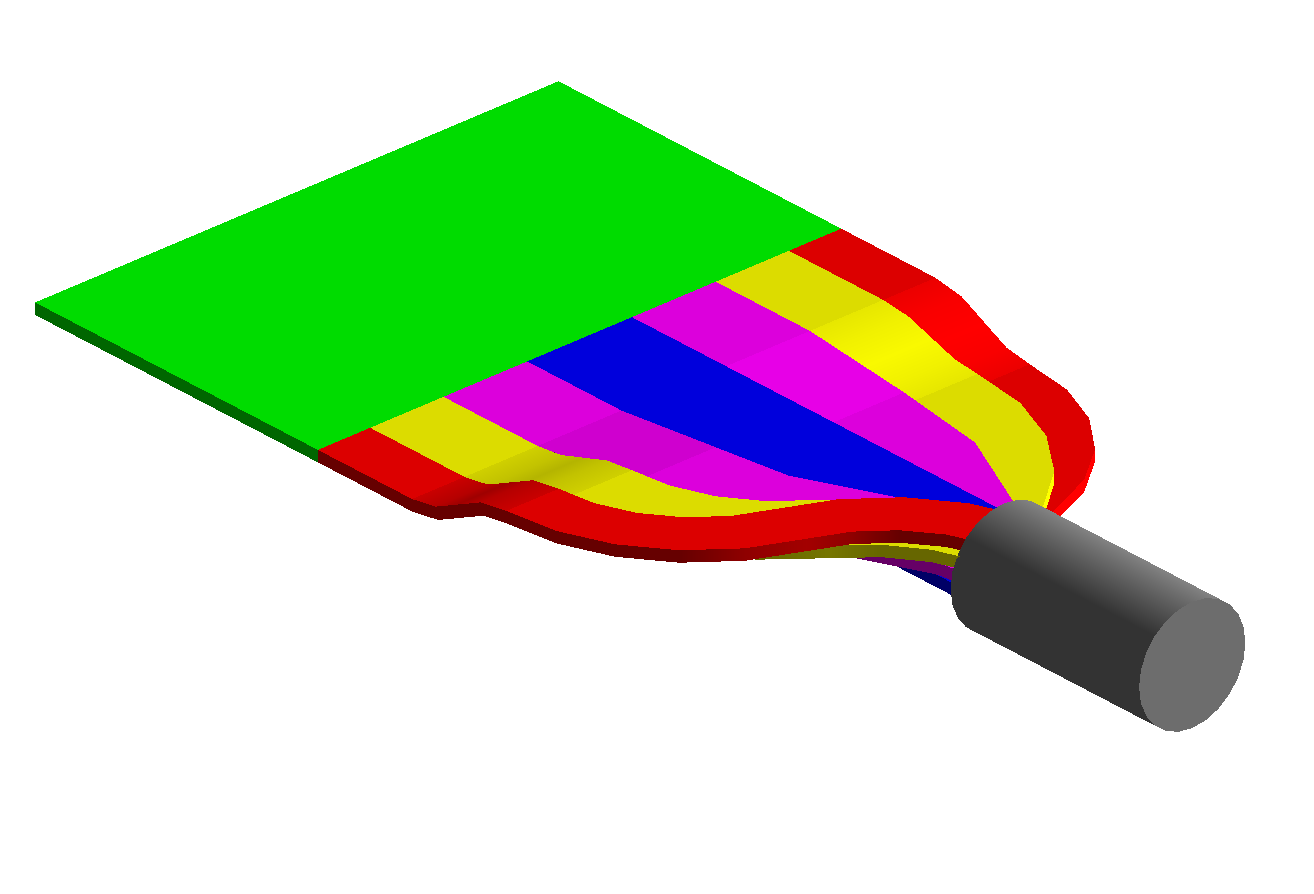}
\hfill
         \includegraphics[trim = 190 -190 0 220, angle = 0, width= 0.25 \linewidth]{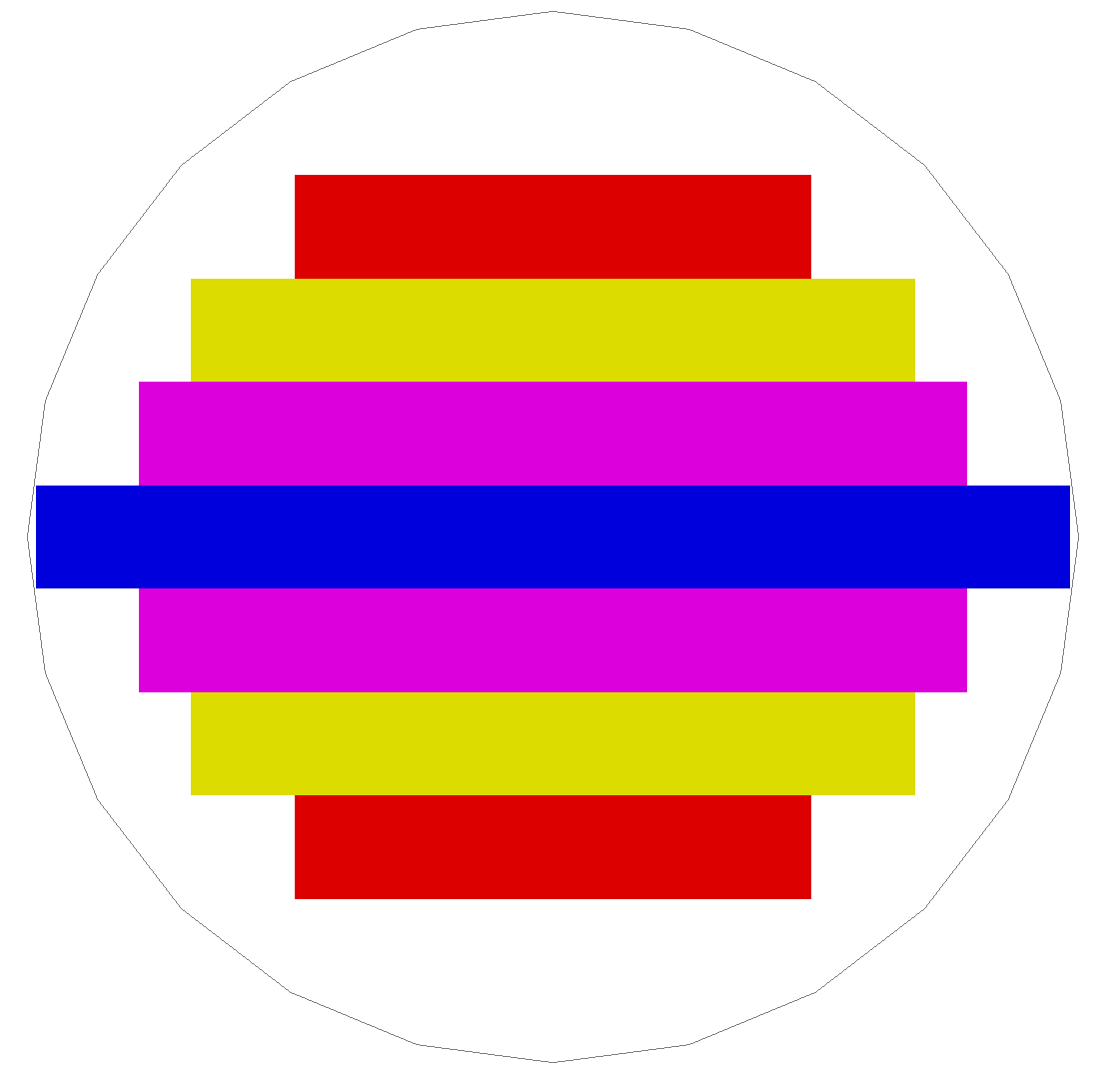}
	\caption{A view of the 7-strip S-shaped LG. 
 The scintillator (15~cm x 25~cm x 0.5~cm) is shown in green.
 The left panel shows the arrangement of the strip at the PMT side of the LG. 
 The width of the light guide is 25~cm. The full length of the LG is 25~cm as it was in the 3-strip LG.}
\label{fig:7strip-model}
\end{figure}

The adiabatic LG between the scintilator with readout area $\bf a \times b$ and the PMT with a diameter $\bf d$ could be designed for $\bf a \times b \leq \pi d^2/4$ (without the use of a light concentrator).
The number of strips in the LG is defined by the ratio of the PMT diameter and the thickness of the scintillator detector.
Naturally, the width of the strip varies, see e.g. Fig.~\ref{fig:7strip-model}.

\begin{figure}[!ht]
	\centering
        \includegraphics[trim = 0 0 0 0, angle = 0, width= 0.8 \linewidth]{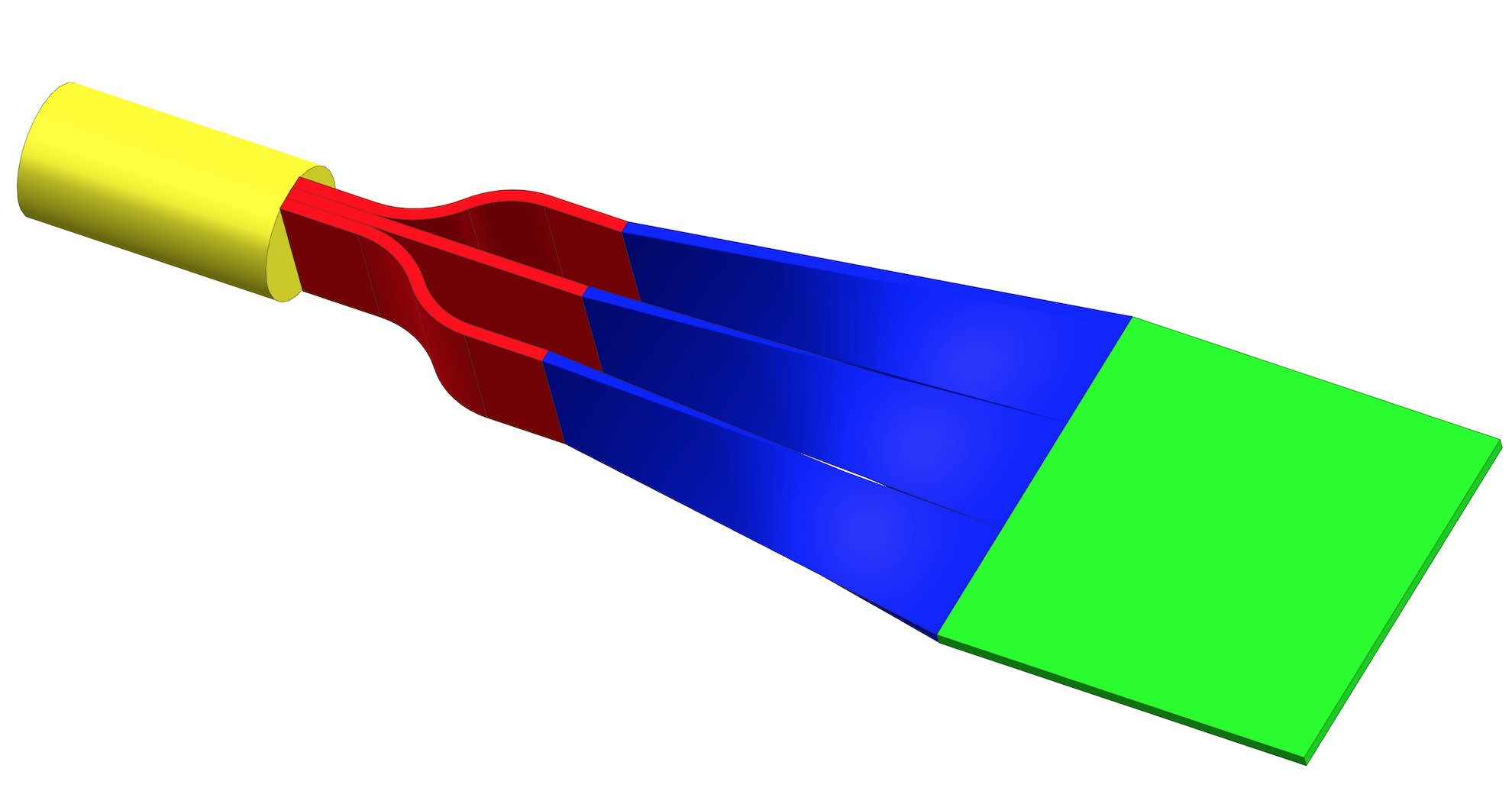}
	\caption{The view of the 90$^o$ twisted LG. 
 The scintillator (15~cm x 15~cm x 0.5~cm) is shown in green. The 5~cm diameter PMT is shown in yellow. 
 The full length of the LG is 25~cm.}
\label{fig:3D-traditional}
\end{figure}

A model of a 3-strip traditional 90$^o$ twisted LG with a scintillator is shown in Fig.~\ref{fig:3D-traditional} and the result for the transmission coefficient is 78.8\%.

\begin{figure}[!ht]
         \includegraphics[trim = 0 120 0 120, angle = 10, width= 0.8 \linewidth]{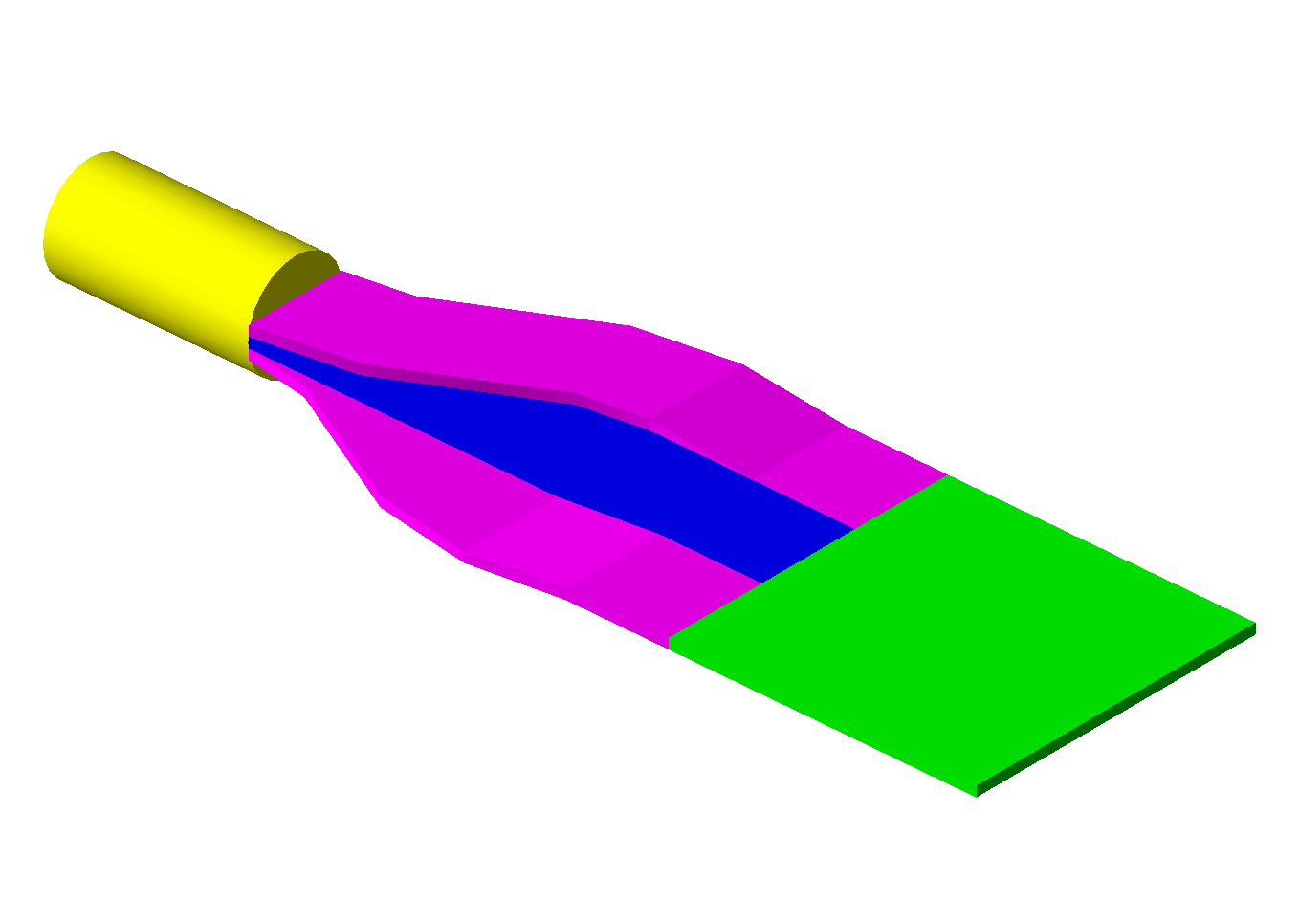}
	\caption{The view of the counter with the 3-strip S-shaped LG. The scintillator 15~cm x 15~cm x 0.5~cm is shown in green. The 5~cm diameter PMT is shown in yellow. The full length of the LG is 25~cm.}
\label{fig:3strip-model}
\end{figure}

A model of a 3-strip S-shaped LG with a scintillator is shown in Fig.~\ref{fig:3strip-model} and the result for the transmission coefficient is 85.3\%.

\begin{figure}[!ht]
	\centering
        \includegraphics[trim = 0 0 0 0, angle = 0, width= 1.0 \linewidth]{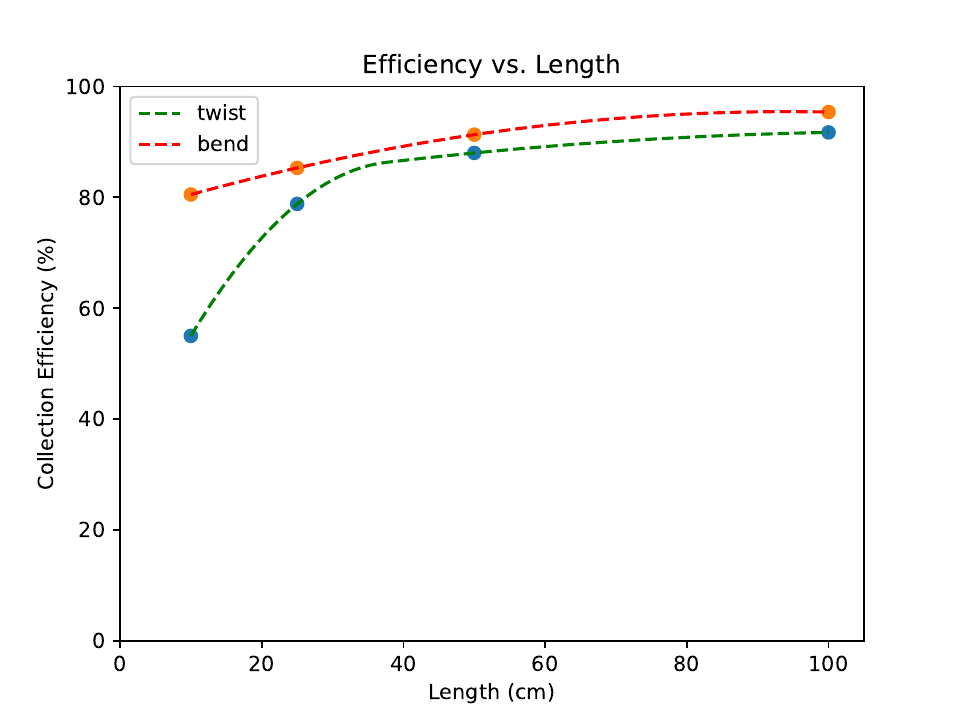}
	\caption{The transmission coefficient vs. length of the light guides.}
\label{fig:eff-length}
\end{figure}

The direct comparison of transmission via the 90$^o$ twisted LG and S-shaped LG shown in Fig.~\ref{fig:eff-length} demonstrates the advantages of the new one in addition to its reduced cost of construction.

Similar results were obtained for a 7-strip option of the LG whose model is shown in Fig.~\ref{fig:7strip-model}.
The light transmission coefficient was found to be 92.2\% for this type of light guide with a 25~cm length (compared to 85.3\% for the same length 3-strip LG).

\section{Spatial uniformity of the light collection}

\begin{figure}[ht!]
        \centering
        \includegraphics[width=0.9 \linewidth]{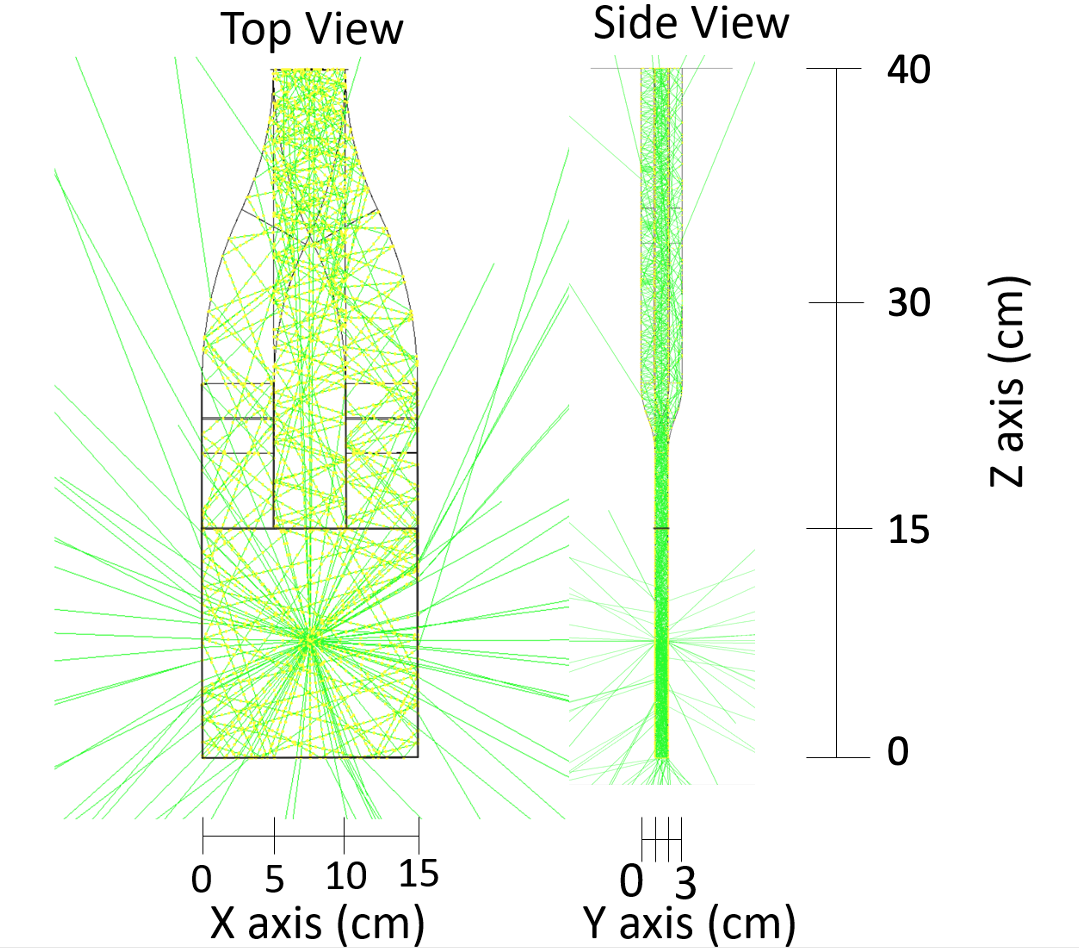}
        \caption{Geant4 MC simulation of light propagation in the 3-strip S-shaped design of the light guide.}
\label{fig:G4MC}
\end{figure}

The spatial variation of the light collection for a 15~cm wide scintillator was investigated in the case of a 25~cm long 3-strip S-shaped LG and a 2" diameter PMT, shown in Fig.~\ref{fig:G4MC}.

\begin{figure}[ht!]
    \centering
    \includegraphics[trim = 0 0 0 0, width=1.0 \linewidth]{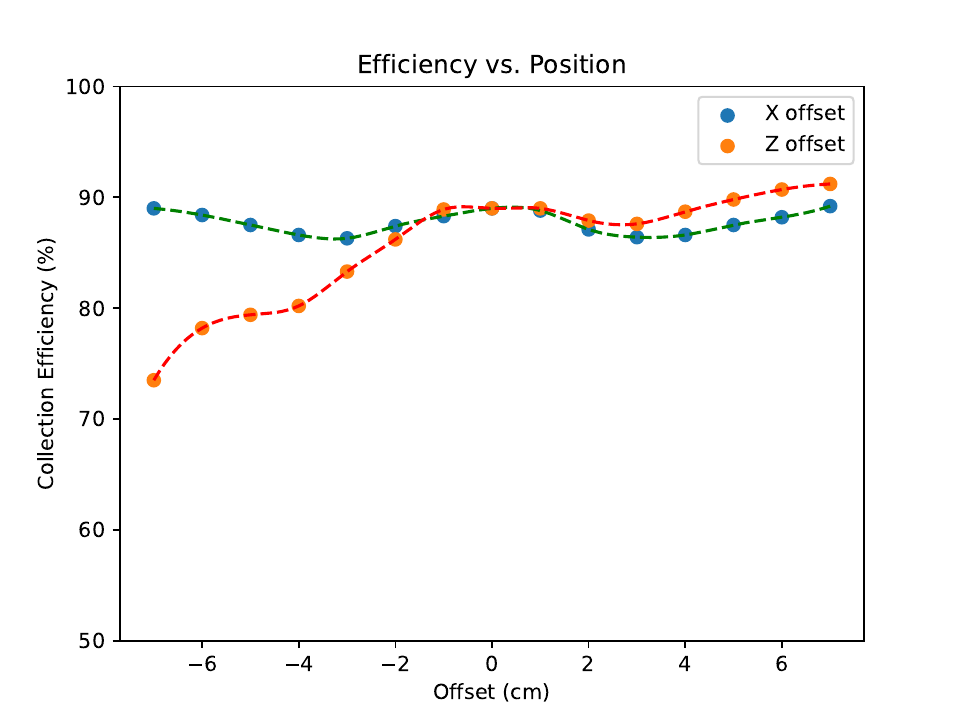}
    \caption{Collection efficiency of the 3-strip light guide vs. position of the light source within the scintillator.
    In scan vs. x the z-position is 7.5 cm and in scan vs. z the x-position is 7.5 cm.}
\label{fig:CE_x-z}
\end{figure}

As can be seen in Fig.~\ref{fig:CE_x-z} the light collection is close to 90\% of the maximum possible with spatial variation on the level of 5\% vs. the x-position, and about 10-15\% for the z-position.

\section{Timing property of the photon propagation}

In this study, we again used a scintillator with a thickness of 0.5~cm and dimensions 15~cm by 15~cm,  Fig.~\ref{fig:G4MC}.

\begin{figure}[ht!]
    \centering
    \includegraphics[width=0.95 \linewidth]{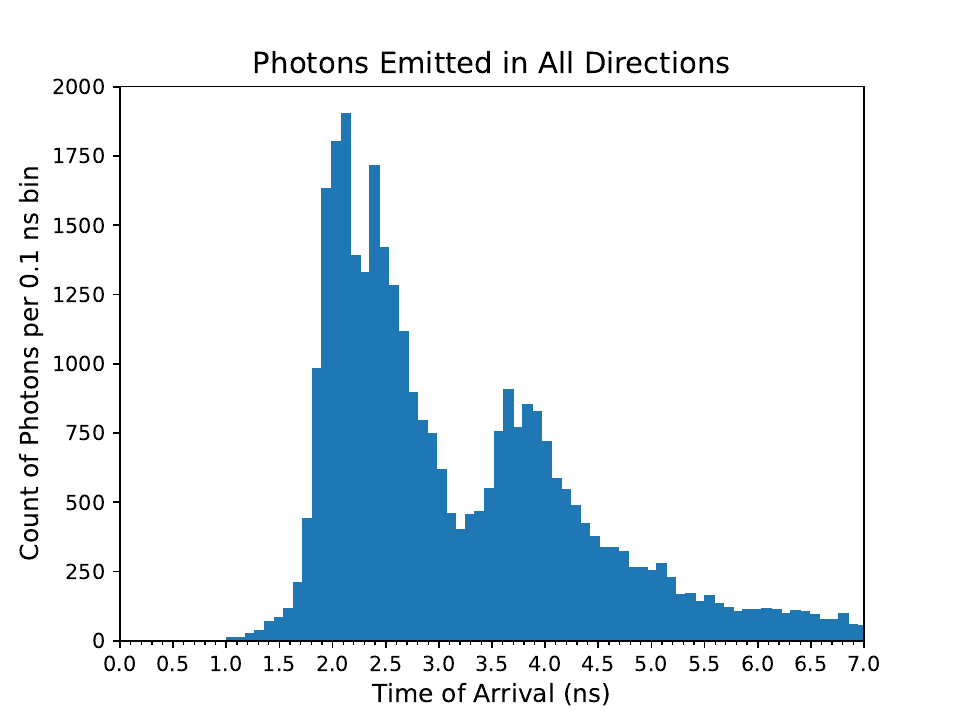}
    \caption{Photon arrival time at PMT for all isotropically emitted photons.}
\label{fig:TOA}
\end{figure}

\begin{figure}[ht!]
    \centering
    \includegraphics[width=0.95 \linewidth]{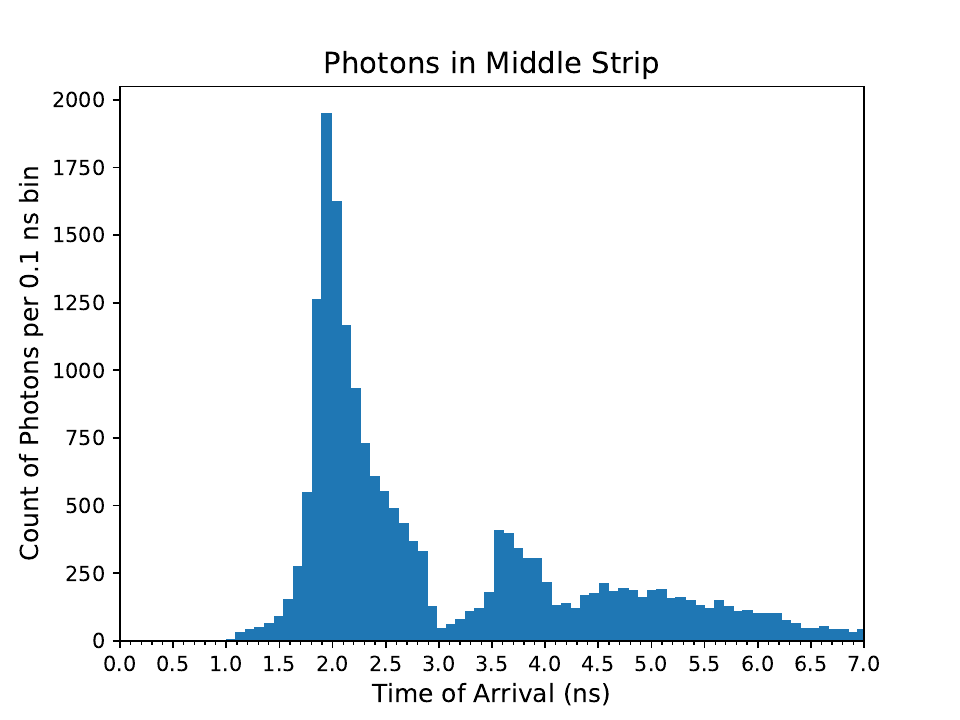}
    \caption{Photon arrival time at PMT for the photons that travel through the central strip.}
\label{fig:TOA_mid}
\end{figure}

The photon time propagation from the source to the PMT was investigated for isotropic emission and also for emission in the direction opposite the LG.
The photon arrival time for the first case is shown in Fig.~\ref{fig:TOA} where there are two peaks, one at 2.2 and another at 3.9~ns.
The first peak consists of a front portion: photons going through the central strip of the LG, and a 0.5~ns delayed portion: those photons going via a number of reflections before entering the LG and having a wider spread of emission angles (see Fig.~\ref{fig:TOA_mid}).

\begin{figure}[ht!]
    \centering
    \includegraphics[width=0.95 \linewidth]{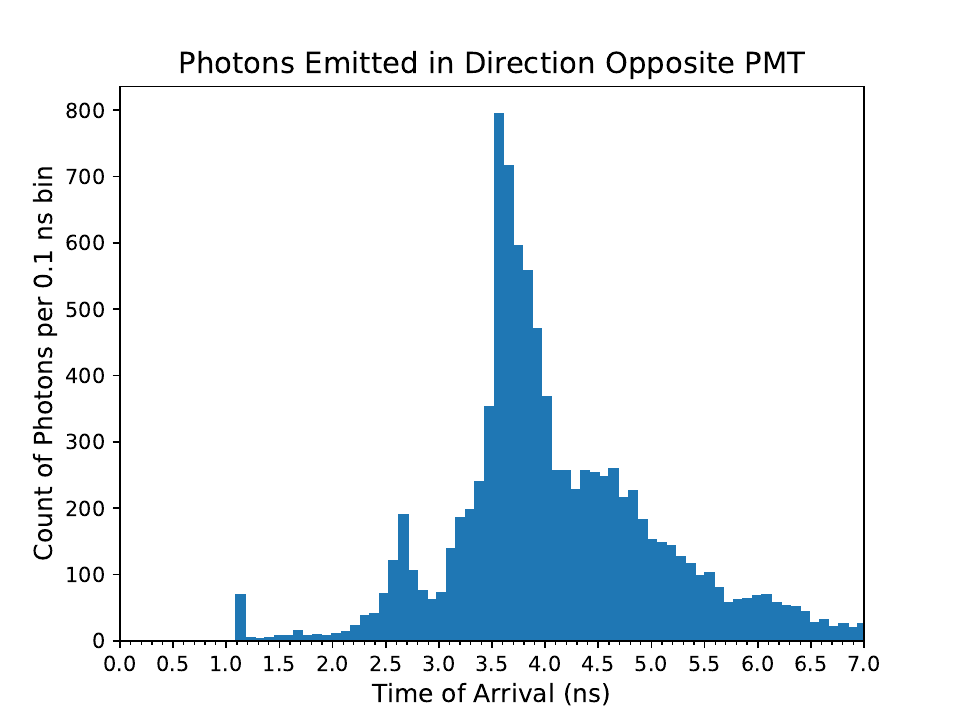}
    \caption{Photon arrival time at PMT for the photons emitted in the direction opposite the LG.}
\label{fig:Top_2nd}
\end{figure}
The event distribution in Fig.~\ref{fig:Top_2nd} confirms that emission in the direction opposite the LG is the origin of the second peak in time of arrival.

\section{Construction methods}

\begin{figure}[ht!]
	\centering
        \includegraphics[width=1. \linewidth]{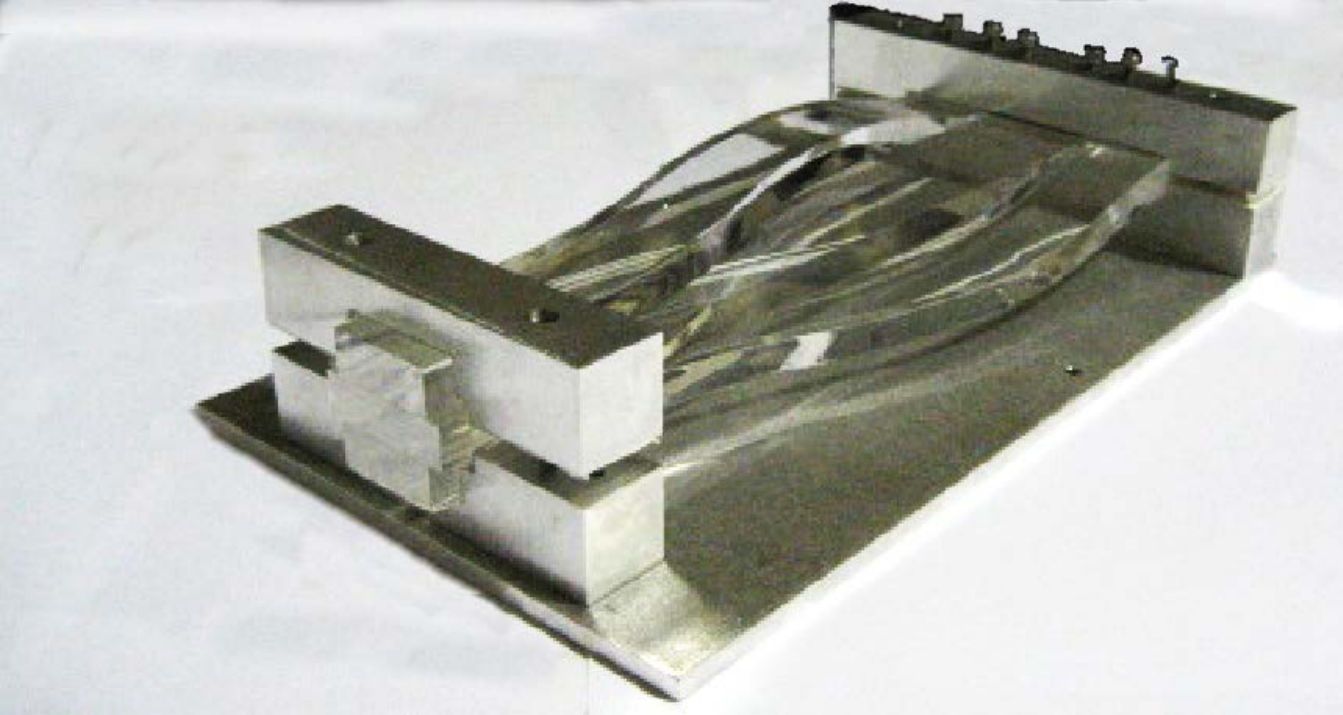}
	\caption{The holder with a light guide used at JLab for LG construction.}
\label{fig:JLab}
\end{figure}

At JLab, for construction of the LG we used the 10 mm thickness S-shaped strips produced by Eljen Technology~\cite{web:ELJEN}. 
An oven was used to do bending of the strips out-of-plane using a jig.
Finally, a holder was used for machining the strip ends, see Fig.~\ref{fig:JLab}.

\begin{figure}[ht!]
	\centering
        \includegraphics[width=1. \linewidth]{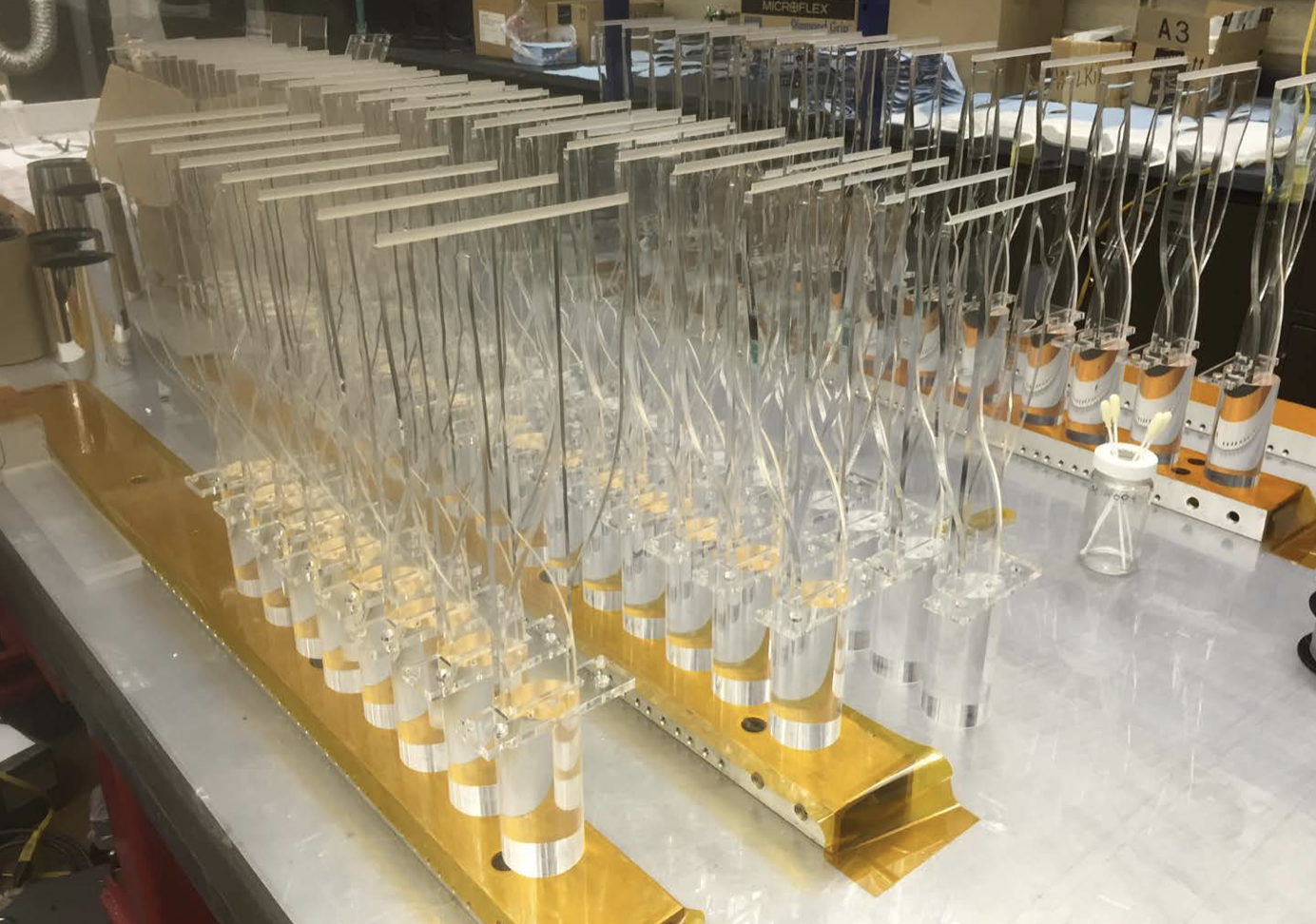}
	\caption{3-strip adiabatic LGs made at CMU.}
\label{fig:CMU}
\end{figure}

At CMU, the 0.5~cm thick S-shaped strips were cut from a sheet of plexiglass by using a commercial laser cutter.
The laser-cut edges were then flame polished using a hydrogen-oxygen torch.
The LG was used for concentration of the light from an area 15~cm by 0.5~cm to the 2" diameter PMT.
Tests showed that this resulted in superior light transmission when compared to acetylene-based flame polishing.
A custom bending fixture was made to heat and bend the strips.
Totally, about 300 LGs were constructed for the segmented hadron calorimeter HCAL-J~\cite{HCAL}, see Fig.~\ref{fig:CMU}.
\vskip 0.15 in

\section{Summary}

The results have shown that the new S-shaped, three-strip light guide configuration has very good light collection efficiency. 
Furthermore, the smaller bending angle has allowed for the possibility of a shorter light guide and an associated variation in the photon time propagation. 
Additionally, the proposed light guide is a cost-effective option for a light guide because it requires much less manpower to construct. 
This concept of a light guide could be useful in solar energy systems as it allows the transfer of the light collected in a large area to a compact photo-detector.

\section{Appendix A: The Geant4 based code}
The code (can be found in Refs.~\cite{Zimmerman,Rosso}) is a modified version of the “OpNovice2” example provided by CERN, with the modifications made to ActionInitialization.cc, DetectorConstruction.cc, EvenAction.cc (added), HistoManager.cc, PrimaryGeneratorAction.cc, Run.cc, RunAction.cc, SteppingAction.cc, and their corresponding header files. 
In ActionInitialization.cc, the event action was added. 
In DetectorConstruction.cc, the physical/logical volumes for all the pieces are constructed and placed accordingly, the materials were given their respective refractive indexes (plexiglass n=1.50, and air n=1.0), and the optical surfaces were defined. 
In EventAction.cc, there are counter variables created to count the photons that reach the different locations in the geometry. 
In HistoManager.cc, there are several histograms added to illustrate results from the simulations, such as time of arrival or initial polar/azimuthal angles of the photons. 
In PrimaryGeneratorAction.cc, extra code was added to emit photons with random azimuthal and polar angles, thus giving them a random direction in three dimensions. 
In Run.cc, extra code was added to title the axes of the added histograms. 
In RunAction.cc, code was added to retrieve the values of the counters that count the photons that reach various volumes and also print those results. 
In SteppingAction.cc, there is an ``if" statement that checks to see if the photon is in the volume of the phototube, and if it is, one is added to a counter, the time of arrival is stored in a histogram and the track of the photon is stopped. 
The counter then gets stored and the final value is sent to the run action where it is printed. 
The stepping action also retrieves the initial polar/azimuthal angles where they are then stored in the histograms in the cases of transmission/reflection. 
For each of the files modified, code was added to their corresponding header files to initialize variables or to return values from one source file to another. 

\section{Acknowledgments}
This work was supported by the CNU graduate program and the US DOE SULI program at Thomas Jefferson National Accelerator Facility. 
Support from P.~Monaghan is appreciated.
We would like to extend our gratitude to A.~Blitstein and G.~Niculescu for assistance with Geant4.
This work was supported in part by the Science Committee of Republic of Armenia under grant 21AG-1C085, the Natural Sciences and Engineering Research Council of Canada and by the US DOE Office of Science and Office of Nuclear Physics under the contracts DE-AC02-05CH11231, DE-AC02-06CH11357, and DE-SC0016577, as well as DOE contract DE-AC05-06OR23177, under which JSA, LLC operates JLab.

\nocite{*}
\bibliographystyle{apsrev4-1}
\bibliography{aipmain.bib}

\end{document}